\documentclass[aps,amsfonts,nofootinbib,superscriptaddress]{revtex4}   
\usepackage{epsfig} 
\usepackage{feynmp}
\usepackage{appendix}
\usepackage{graphicx}
\usepackage{dsfont}
\usepackage{amssymb}
\usepackage{tipa}
\usepackage{amsmath}
\usepackage{amsbsy}
\usepackage{color}
\usepackage{slashed}
\usepackage{bm}
\usepackage{bbm}
\usepackage{float}
\usepackage{ulem}
\usepackage{tikz}
\usepackage{pgfplots}
\usepackage{xparse}
\usepackage{upgreek}
\usepackage{stmaryrd }
\usepackage{caption}
\usepackage{subfigure}
\usepackage{soul}
\usepackage{dutchcal}

\usetikzlibrary{positioning,arrows,patterns,automata}
\usetikzlibrary{decorations}
\usetikzlibrary{trees}
\usetikzlibrary{decorations.pathmorphing}
\usetikzlibrary{decorations.markings}
\usetikzlibrary{calc}

\newcommand{\ee}{\end{equation}}
\newcommand{\bb}{\begin{equation}}
\newcommand{\eqb}{\begin{eqnarray}}
\newcommand{\eqf}{\end{eqnarray}}

\def\sigmavec{\mbox{\boldmath$\sigma$}}
\def\sigmavec{\mbox{\boldmath$\sigma$}}

\def\ca{{\mathbcal a}}
\def\cb{{\mathbcal b}}
\def\cc{{\mathbcal c}}
\def\cd{{\mathbcal d}}
\def\ca{{\mathbcal a}}
\def\cb{{\mathbcal b}}
\def\cc{{\mathbcal c}}
\def\cd{{\mathbcal d}}
\tikzset{	aphoton/.style={decorate, decoration={snake}, draw=blue},
	photon/.style={decorate, decoration={snake}, draw=black},
	particle/.style={draw=black, postaction={decorate},
		decoration={markings,mark=at position .5 with {\arrow[draw=black]{>}}}},
	gluon/.style={decorate, draw=red,
		decoration={coil,amplitude=4pt, segment length=5pt}},
	vertex/.style={draw,shape=circle,fill=black,minimum size=3pt,inner sep=0pt},
}
\NewDocumentCommand\semiloop{O{black}mmmO{}O{above}}
{% 
\draw[#1] let \p1 = ($(#3)-(#2)$) in (#3) arc (#4:({#4+180}):({0.5*veclen(\x1,\y1)})node[midway, #6] {#5};)}

\begin{document}
\title{Testing Dark Matter with the Anomalous Magnetic Moment in a Dark Matter Quantum Electrodynamics Model}
 \author{Ashok K. Das}
 \email{das@pas.rochester.edu}
 \affiliation{  Department of  Physics  and  Astronomy, University  of
   Rochester, Rochester, NY 14627-0171, USA}
\affiliation{  Institute  of  Physics, Sachivalaya  Marg,  Bhubaneswar
  751005, India} 
\author{Jorge Gamboa}
\email{jorge.gamboa@usach.cl}
\affiliation{Departamento de  F\'{\i}sica, Universidad de  Santiago de
  Chile, Casilla 307, Santiago, Chile}
\author{Fernando M\'endez}
 \email{fernando.mendez@usach.cl}
\affiliation{Departamento de  F\'{\i}sica, Universidad de  Santiago de
  Chile, Casilla 307, Santiago, Chile}
  \author{Natalia Tapia}
\email{natalia.tapiaa@usach.cl}
\affiliation{Departamento de  F\'{\i}sica, Universidad de  Santiago de
  Chile, Casilla 307, Santiago, Chile}
\date{\today}

\begin{abstract}
  We consider  a model of dark quantum  electrodynamics which  is coupled
  to a visible photon through  a kinetic mixing term. We compute  the $g_\chi-2$ for the dark fermion, where  $g_\chi$ is
  its gyromagnetic factor.  We show that the  $g_\chi-2$ of the dark fermion is related to the $g'_{\chi}-2$
  of  (visible) quantum  electrodynamics through a constant which depends on the  kinetic  mixing
  factor.   We determine $g_\chi-2$ as a function  of   the   mass ratio
  $\kappa  =m_B/m_\chi$ where  $m_B$  and  $m_\chi$ denote  the
  masses of the  dark photon and the dark fermion respectively and we show how $g_{\chi}-2$ become very different for  light and heavy fermions around $m_B  \leq10^{-4}\,$ eV.  
\end{abstract}

\maketitle

\section{Introduction}
The search for dark matter is one of the important challenges in 
physics today  because its existence  may  explain some of the puzzles  of
conventional physics, such as the rotation curves of the galaxies, the new
phenomenon of emission of light from the center of galaxies. It can also provide
new  insights into old  problems such  as matter-antimatter  asymmetry,
primordial magnetic  fields and  so on. Dark  matter can also open up new areas of research  
in particle physics,  astrophysics and cosmology \cite{zeldovich,kolb,review,review11,review12,review13}.

The dark matter interacts very weakly with conventional matter and the effects produced by either light or heavy dark fermions are small in general. On the other hand, some properties, such as the magnetic dipole  moment for electron/muon like dark fermions may depend dramatically on  the mass of the dark fermion as well as the way they couple to the visible sector. (The Pauli coupling, for example, depends on the fermion mass with the mass in the denominator.) The possibility of a mass dependence for  such effects makes the heavy and the light fermion regimes very different since they correspond to different parameter spaces. Precisely for this reason, the distinction between weakly interacting slightly particle (WISP) and weakly interacting massive particle (WIMP) is introduced in these studies.
 
 Although the observational and theoretical reasons for the existence of 
dark matter  are many (for a review see \cite{bertone,Djouadi}), there are 
issues that still need to be addressed. For example, for models mirroring the visible sector, such as the one 
considered in the present work, it would be interesting to find the effects of such dark systems on some of
the high-precision measurements in particle physics \cite{test,olive}.

The electron gyromagnetic factor is one of the greatest triumphs of 
quantum field theory (quantum electrodynamics) and, therefore, the study of the magnetic moment of a dark Dirac fermion would be a very relevant and important issue in this context. However contrary of the conventional standard model calculations, the calculation for the magnetic moment of dark fermions must take into account the range of masses in which the theory is considered. For the case of WISP the typical fermion masses are $m_{wisp} \sim10^{-3}$ eV or less, while in the case of WIMP the fermion masses are typically $m_{wimp} \sim10^2$ GeV or more and, therefore the magnetic moment of a WISP and a WIMP can be very different. 

In this paper we would like to concentrate mainly on the sector of WISP where the space of parameters can be tested in a family of experiments running at present \cite{babb,ry}. The masses of dark light fermions are bounded typically as $m_\chi <10^{-3}$ eV  which can be easily studied in these experiments.

In the low energy sector, which is the region to be discussed in this paper, there are still a number of important phenomena that require careful study such as the distortion of cosmic microwave radiation \cite{ring}, deviations from Coulomb's law \cite{coulomb1,coulomb2}, the distortion of planetary magnetic fields \cite{planet}, the level shifts in atomic physics \cite{shift2} and so on. The ranges of parameters that may be relevant in these studies are $ m_{\chi} \sim 10^{-8} -10^6$ eV for dark fermion masses and $ \xi \sim 10^{-14} -10^{-4}$ for the kinetic mixing parameter (to be introduced in the next section) and, as we will see in the following, this can be perfectly tested with the current experiments mentioned above.

The calculation of the magnetic moment for the light dark fermions \cite{barger,pospelov,kim,kim1,pospelov2} requires particular attention  because of the smallness of the mass of the fermion. In this case,  the loop corrections of the standard model ({\it e.g.} from $Z$ boson loops) are not relevant. In the opposite case of heavy dark fermions, on the other hand, the calculation of the magnetic moment is similar to that of the muon.
 
The paper is organized as follows. In section {\bf II}, we set up the problem and present the model to be studied. Details of the calculations are given in section {\bf III} where we also discuss the implications of the results. In section {\bf IV} we summarize and discuss some open problems and in the appendix we provide some details of the diagonalisation of the mass matrix.
 
\section{Dark Matter and Kinetic Mixing}

In order to understand this problem systematically, let us build our model of {\it dark QED}. 
First, we consider a dark fermion $\chi$
with mass $m_\chi$ coupled  to a hidden (dark) photon $B_\mu$.  We choose the
coupling constant (for minimal coupling) of the dark fermion to  the hidden photon to be unity 
$e_h=1$ for simplicity so that  the covariant derivative in  the dark sector can be written as 
$D_\mu[B]=  \partial_\mu +i  B_\mu$.  We note that since we are interested in 
 dipole interactions, we have to couple the dark fermion non-minimally and we choose the Lagrangian density for the dark fermion to be given by 
\bb 
{\cal  L}_{\chi}   =  {\bar   \chi}  \left(   {{  i   D\hspace{-.6em}  \slash
      \hspace{.15em}}}  [B]  +  \frac{g_\chi}{ \Lambda}\sigma_{\mu  \nu}
  F^{\mu \nu} (B) - m_\chi \right) \chi.
 \label{dark10}
\ee

The Pauli coupling term in the Lagrangian density may arise either from radiative corrections or can be included at the tree level Lagrangian density in an effective theory. Here $\Lambda$ is a mass  scale which, in the case of (visible)  QED, is proportional to  the  electron mass $m$.  Therefore, by analogy,  $\Lambda$ 
can be chosen in the present case to be proportional to  the mass of the dark fermion ($m_\chi$).
The constant $g_\chi$ can be thought of as the gyromagnetic factor for the dark fermion in analogy with standard QED.
 We point out that the ($4\times 4$) matrix $\sigma_{\mu \nu} F^{\mu 
 \nu}$ in the Pauli term can be written explicitly in terms of electric and magnetic fields as 
\bb 
\frac{1}{2}\sigma_{\mu \nu} F^{\mu \nu}  = - \mbox{diag} \left\{ \sigmavec \cdot({\bf
    B}_{h} + i {\bf E}_{h}) \,, \, \sigmavec\cdot ({\bf B}_{h} - i {\bf E}_{h})\right\}, 
\ee 
where ${\bf E}_{h}, {\bf B}_{h}$ correspond respectively to the dark electric and magnetic fields associated with $B_{\mu}$. Therefore  the Dirac fermion, coupled non-minimally to the gauge  field $B_\mu$, 
contains both  an electric  dipole  term  $\sigmavec\cdot {\bf  E}_{h}$  and  a
magnetic  dipole term $\sigmavec\cdot {\bf B}_{h}$. We note here that  in spite of the eventual importance of magnetic moment in the search for dark matter, at present there is no discussion of this in the literature beyond the tree level (except for \cite{pospelov}, \cite{foadi} within the context of technicolor dark matter and an interesting  recent reference  \cite{paolo} within the context of early universe). In this work, we carry out the analysis of the magnetic moment at the one loop level.

Next, we choose the dynamics for the hidden photon to be given by the Lagrangian density
\bb
{\cal  L}_B  =  - \frac{1}{4} F_{\mu \nu} (B) F^{\mu
  \nu} (B) + \frac{m_{B}^2}{2}B_\mu B^\mu. 
\label{dark1} 
\ee  
Here $m_B$  is the mass  of  the  dark photon which can arise, for example, through the Higgs mechanism or the 
St\"uckelberg formalism in the hidden sector. 
Similarly, in the visible sector, the dynamics of the photon can be given by the standard Maxwell term. We note that although the mass of the hidden (dark) photon $m_B$ may be very small, we find from our analysis that the relevant parameter in the study of the magnetic moment is the ratio of the masses of the dark photon and the dark fermion (which we denote later as $\kappa = m_B/ m_\chi$). This ratio can be extracted appropriately from the experimental data and, indeed, can give nontrivial effects even when the individual masses are small (see Figures 2 and 3 later).

We will assume the kinetic mixing model of \cite{Holdom} which allows for a mixing between the gauge bosons in the visible and the hidden sectors through the Lagrangian density
\begin{equation}
{\cal L} = - \frac{1}{4}\, F_{\mu\nu}(A) F^{\mu\nu} (A) + \frac{m_A^2}{2} A_\mu A^\mu + \frac{\xi}{2}\, F_{\mu\nu}(A) F^{\mu\nu} (B) + {\cal L}_B.\label{dark2}
\end{equation}
Here $\xi$ is a dimensionless mixing parameter assumed to be small and  we can also add a term which represents the interaction of the photon with the visible charged current \cite{engel}. However, we neglect this term for simplicity since it is not relevant for our analysis.  In the present model  we have admitted a mass 
$m_A$ for the photon of the visible sector. This mass is presently constrained by the fact that there exist galactic magnetic fields coherent on the galactic size of a few kpc \cite{pdg,dolgov} leading to
$$m_A < 1\times 10^{-27}\, \mbox{eV}, ~~~~~\mbox{or,} ~~~~~~~~~ \lambda_\gamma >10^{22} \mbox{cm},
$$ 
where $\lambda_{\gamma}$ denotes the Compton wave length of the photon.

The simultaneous  diagonalisation of the kinetic mixing and the mass terms  is achieved in two steps (for 
details see the appendix) allowing us to write the (diagonalized) Lagrangian density
\begin{equation}
{\cal L} =   -  \frac{1}{4} F^2 (A^{''}) -\frac{1}{4}F^2(B^{''}) + \frac{m_B^2}{2}  (B^{''})^2  + \frac{m_A^2}{2} (A^{''})^2
 + {\bar  \chi}   \left({{   i\partial\hspace{-.6em}\slash
       \hspace{.15em}}}      +  \xi_B   {{B\hspace{-.6em}\slash
       \hspace{.15em}}}^{''}   +   {\xi}_A {{A\hspace{-.6em}   \slash
       \hspace{.15em}}}^{''} - m_\chi \right) \chi + {\cal L}_{\sc P}, \label{dark3}
       \ee
where  $\xi_A\approx -({m_A}^2/m_B^2)\xi$ and ${\xi}_B\approx 1$  for $\xi \ll1$, which is the 
case we study in our model. With these redefinitions, the Pauli term  ${\cal L}_{\sc P}$ takes the form 
\bb
 {\cal L}_{\sc P} = \frac{g_\chi}{\Lambda}\, {\bar  \chi} \left(\sigma_{\mu
    \nu} F^{\mu \nu} (B^{''}) + \xi_A \sigma_{\mu \nu}
  F^{\mu \nu} (A^{''}) \right) \chi.
 \label{dark4} 
\ee 
We see that the diagonalization introduces a coupling of the dark fermion to the visible photon both in the minimal as well as in the nonminimal terms.

For $\Lambda=8 m_\chi$, we see from  (\ref{dark4}) that $g_\chi$  in the Pauli interaction 
can be thought of as the gyromagnetic factor of the dark fermion whose
magnetic moment is given by (${\bf S} = \frac{1}{2} \sigmavec$ denotes the spin of the fermion)
$$
\mbox{{\boldmath$\mu$}}\equiv-\frac{g_\chi}{2m_\chi}{\bf S},
$$
in complete analogy with standard QED. However, the second term in \eqref{dark4}
\[
 \frac{ g_\chi\,\xi_A}{8 m_\chi}{\bar \chi} \sigma_{\mu \nu}  F^{\mu\nu} (A^{''}) 
 \, \chi, 
 \]
represents an interaction of the dark fermion with the visible photon.  In the static limit of a constant
(visible) magnetic field ${\bf B}$ (${\bf B}$ is the magnetic field associated with the visible photon $A_{\mu}^{''}$ and is not the space component of the dark photon $B_{\mu}^{''}$) this term leads to
 \[
 \frac{ g_\chi\,\xi_A}{8 m_\chi}{\bar \chi} \sigma_{\mu \nu}  F^{\mu\nu} (A^{''}) 
 \, \chi \xrightarrow{\sc{{\rm static}\, \mathbf{A^{''}}}}{\bar \chi}  \bigg[
\left( -\frac{ g'_\chi\,}{2 m_\chi}  {\bf S}\right)\cdot {\bf B}
 \, \bigg] \chi.
 \]
Therefore, the  coefficient $g_\chi'\equiv g_\chi\,\xi_A$ plays the role of the gyromagnetic factor for the
interaction of the dark fermion with the visible photon. This is interesting because although $g_\chi$ is 
an unknown parameter, as  it originates in processes in the dark sector of the 
theory (namely, the first term in the Pauli coupling in \eqref{dark4}), the second term involves the interaction with the visible photon opening up new possibilities for its measurement. 

The one loop quantum vertex corrections arising  
from both the minimal coupling terms in \eqref{dark3} involve the two diagrams shown in Figure \ref{fig1}. Fig. \ref{figa} denotes the vertex diagram contributing to the gyromagnetic factor in the dark 
sector, while Fig. \ref{figb} contributes to the gyromagnetic factor for the interaction of the dark fermion 
with the photon in the visible sector. The detailed calculations for these are described in the following section.

\section{Details  of the Calculation}
  
We have already seen that in our model we can identify two gyromagnetic
factors. One of them ($g_\chi$) is a  genuine   correction to the magnetic 
moment  of the dark fermion obtained from diagram Fig. \ref{figa},
while the other comes from the interaction with the visible photon, as 
is shown in Fig. \ref{figb}. The integrands for the two diagrams  are the same (except for a multiplicative constant)  and, therefore, the calculations can be done in the conventional manner \cite{schwinger}.  Even though this calculation
 is well known, we will describe this in some detail in order to clarify 
 differences with the standard case.  
 
 The effect of loop contributions to the vertex is equivalent to modifying the vertex $
 \gamma^\mu \to \Gamma^\mu$ where the new vertex function can be parameterized as  
 \bb 
\Gamma^\mu = \gamma^\mu\, F_1\left(\frac{p^2}{m_\chi^2}\right) + \frac{\sigma^{\mu \nu}}{2 m_\chi} p_\nu\,F_2\left(\frac{p^2}{m_\chi^2}\right).\label{vertex}
\ee
Here,  $F_1$ and $F_2 $ are two form factors which depend on the
Lorentz invariant combination $(p^2/m_\chi^2)$ of the  transferred 
momentum $p_\mu$ and   the mass scale of the fermion $m_\chi$  . 
Therefore, the diagram in Fig. \ref{figa} can be thought of as giving rise to the amplitude
\bb 
i {\cal M}^{\mu} = -i\bar{u}(q_{2})\Gamma^{\mu} u(q_{1}) = -i {\bar u}(q_2) \left[ F_1\left(\frac{p^2}{m_\chi^2}\right) \gamma^\mu+ \frac{\sigma^{\mu \nu}}{2 m_\chi} p_\nu F_2\left(\frac{p^2}{m_\chi^2}\right) \right] u(q_1), 
\label{cal2}
\ee
where $q^{\mu}_1, q^{\mu}_{2}$ are the four momenta of the incoming and 
the outgoing fermions respectively. The four momentum of the external photon
defines the momentum transfer $p^{\mu} =q^{\mu}_2-q^{\mu}_1$. Diagram in Fig. \ref{figb} gives rise to a similar contribution except that  
the coupling of the dark fermion with visible photon  has an extra factor of $\xi_A$ which can be seen from the Lagrangian density (\ref{dark3}). The two form factors appearing in (\ref{cal2}) (or \eqref{vertex})  have well known physical interpretations.  
While $F_1$ contributes to charge renormalization,  
$F_{2}$ provides a genuine contribution to the magnetic moment of the fermion. In fact,
the gyromagnetic factor $g$ can be written in terms of $F_{2}$ as   
\bb
g = 2(1 +  F_2 (0)).\label{g}
\ee

%%%%% FIGURE 1%%%%%%%%%%%%%%%%%
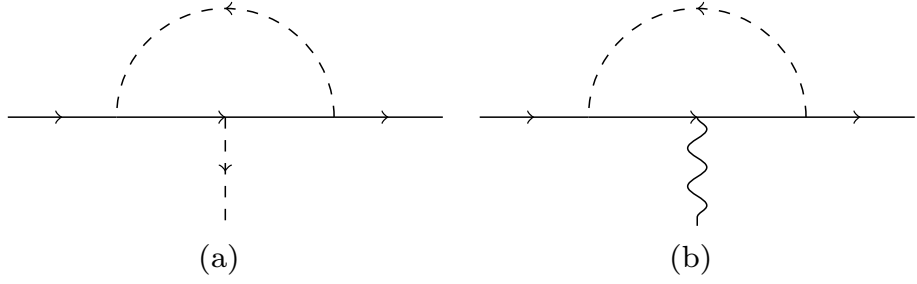
\begin{figure}[t!]
\centering
\begin{minipage}[c]{0.7\textwidth}
  \resizebox{\textwidth}{!}{
\subfigure [\label{figa}] {
\begin{tikzpicture}[node distance=1cm and 1cm]
\coordinate (v);
\coordinate[right=of v] (v1);
\coordinate[right=of v1] (v2);
\coordinate[left=of v](v3);
\coordinate[left=of v3](v4);
\coordinate[below=of v](v5);
\coordinate[below=of v5](v6);
\draw[particle] (v4) -- (v3);
\draw[particle] (v3) -- (v1);
\draw[particle] (v1) -- (v2);
\semiloop[dashed, particle]{v3}{v1}{0};
\draw[dashed, particle] (v) -- (v5);
\end{tikzpicture}
}
\subfigure [\label{figb}] {
\begin{tikzpicture}[node distance=1cm and 1cm]
\coordinate (v);
\coordinate[right=of v] (v1);
\coordinate[right=of v1] (v2);
\coordinate[left=of v](v3);
\coordinate[left=of v3](v4);
\coordinate[below=of v](v5);
\coordinate[below=of v5](v6);
\draw[particle] (v4) -- (v3);
\draw[particle] (v3) -- (v1);
\draw[particle] (v1) -- (v2);
\semiloop[dashed, particle]{v3}{v1}{0};
\draw[photon] (v) -- (v5);
\end{tikzpicture}
}}
\captionsetup{font=footnotesize}
\captionof{figure}{\label{fig1}The two possible vertex corrections involved 
in the gyromagnetic factor of dark matter (a dashed line represents a hidden 
photon propagator). The diagram (a) is  the  genuine 
contribution to the magnetic moment coming from the coupling of the dark fermion
to the hidden photon. Diagram (b), on the other hand, corresponds to the contribution coming from the coupling of the dark  fermion to the visible photon. Both diagrams differ by a factor of $\xi_{A}$.}
\end{minipage}
\end{figure}
%%%%%%%%% END FIGURE 1%%%%%%%%%%%%%%%%

Form factors can be calculated from the diagrams as follows. The amplitude 
for the diagram in Fig. \ref{figa} is given by   
\bb
i{\cal M}^\mu = -\, {\bar u}(q_2) \int \frac{d^4k}{(2\pi)^4} 
\frac{\eta_{\alpha\beta} 
\gamma^\alpha ({{  p\hspace{-.6em}  \slash \hspace{.15em}}} +
{{ k \hspace{-.6em}  \slash \hspace{.15em}}} + m_\chi) 
\gamma^\mu 
({{ k\hspace{-.6em}  \slash \hspace{.15em} + m_\chi)
\gamma^\beta}}}{[(k-q_1)^2 - m_ \gamma^2+ i\varepsilon] 
[(p+k)^2 - m_\chi^2 + i\varepsilon] [k^2 -m_\chi^2] + i\varepsilon]}
u(q_1).
\ee
After some algebra (long but straightforward)  one can identify contributions proportional to 
the matrix $\sigma^{\mu\nu}p_\nu$ which we are interested in  
(see  (\ref{cal2})), to correspond to ($x,y,z$ denote Feynman combination parameters)
\bb 
F^{(a)}_2 (p^2) = - i\,8\,m_\chi^2  \int_0^1 dx dy dz \delta (x+y+z-1) \int \frac{d^4k}{(2\pi)^4} \frac{z(1-z)}{(k^2-\Delta + i\varepsilon)^3} + \cdots 
\ee
where $\Delta = -x y p^2 + (1-z)^2 m_\chi^2 + m_B^2 z$. 

The integral over $k$ can be calculated in the standard manner \cite{schwinger} and leads to 
\eqb 
F^{(a)}_2 (p^2) &=& \frac{1}{8 \pi^2} \,\int_0^1 dx\, dy\, dz \, \delta(x+y+z-1) \frac{z(1-z)}{(1-z)^2  
+ \kappa^2 z - x\,y \,\frac{p^2}{m_\chi^2}}.
\eqf
with $\kappa^2 =m_B^2/m_\chi^2$. This is the standard result of QED 
if the photon is massive which is the case for our hidden photon. 
 
In the limit  $p_\mu \to 0$, the form factor turns out to be 
\bb 
F^{(a)}_2 (0) = \frac{1}{8 \pi^2} \int_0^1 dx\, dy\, dz \, \delta(x+y+z-1) \frac{z(1-z)}{(1-z)^2  + \kappa^2 z }. 
\label{fmass}
\ee 
Furthermore, in the limit $\kappa \to 0$ (massless dark photon), the value of the integral is $1/2$ and, in this case, 
$2F_2(0)=\alpha/2\pi$ (if we reinsert the factors of charge $e^2$ which we 
have set to unity in the hidden sector) which coincides with Schwinger's result for the 
gyromagnetic factor  $g-2$ for QED \cite{test}. 
Furthermore, for diagram \ref{figb}, which gives rise  to the magnetic moment of 
the dark fermion through coupling to the visible photon, the calculation is completely parallel and we can write
the results for the contributions from the two diagrams (to the magnetic moment) as
\begin{equation}
F^{(a)}_2 (0) = \frac{1}{8\pi^2}  f(\kappa), \quad
F^{(b)}_2 (0) = \frac{\xi_A}{8 \pi^2} f(\kappa). \label{F}
\end{equation} 
  
The function $f(\kappa)$ can be obtained from (\ref{fmass}) and a direct calculation
gives 
\bb 
f(\kappa) = -\frac{ \kappa \left( 2- 4\kappa^2 + \kappa^4\right)}{\sqrt{4-\kappa^2}} \,
\tan^{-1} \left( \frac{\sqrt{4-\kappa^2}}{\kappa}\right) 
+ \frac{1}{2} \left[ 1- 2\kappa^2 +2 \kappa^2 (\kappa^2 -2) \ln \kappa \right].
 \label{f}
\ee 
It may appear that this function is defined only for $0<\kappa<2$. However, this result
is, in fact, valid for all values of $\kappa>0$. This can be seen directly by noting that the imaginary parts 
arising from $\tan^{-1}$ (arctan) cancel with those coming from the square root 
in the denominator. Figure \ref{fig2a} illustrates this (reality) behavior.

%%%%%%FIGURE 2%%%%%%%%%%%%%
\begin{figure}[t!]
\centering
\begin{minipage}[c]{0.4\textwidth}
\subfigure[\label{2a}]{
\includegraphics[scale=0.56]{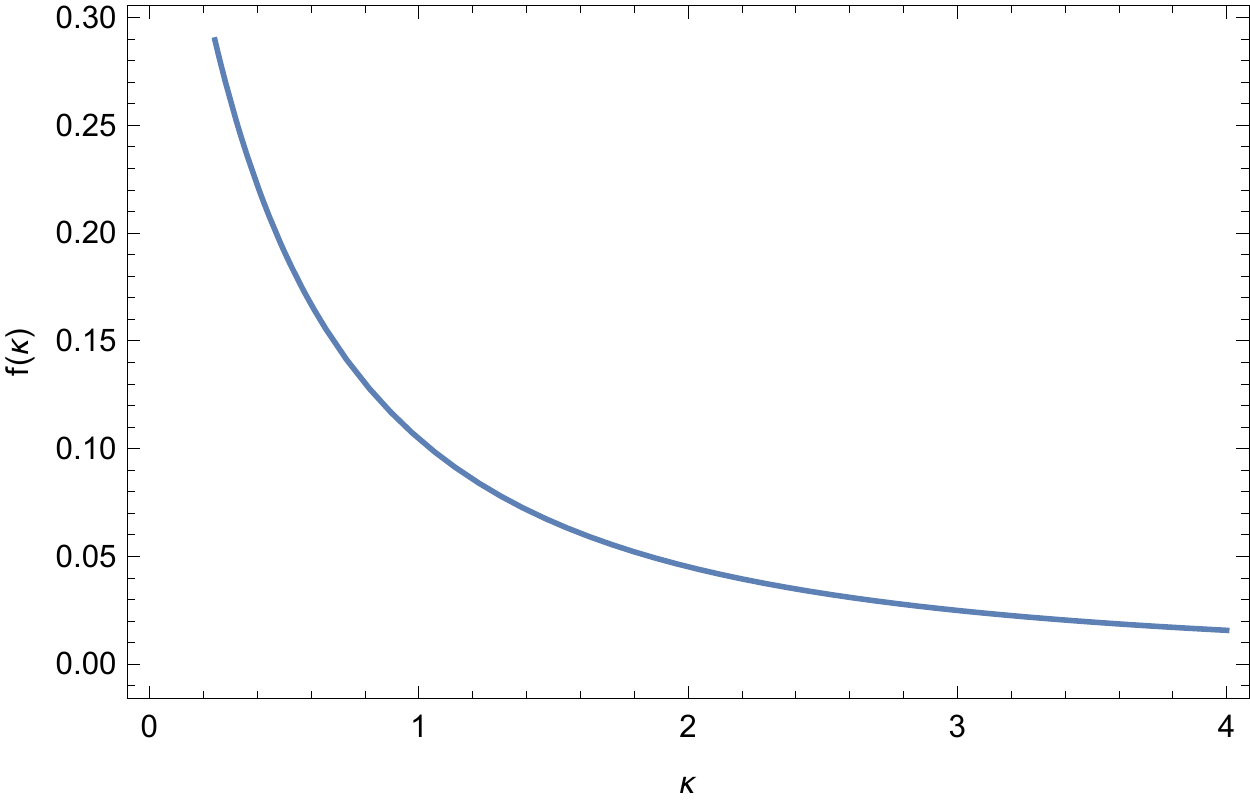}  
\label{fig2a}
}
\end{minipage}
\begin{minipage}[c]{0.4\textwidth}
\subfigure[\label{2b}]{
\includegraphics[scale=0.56]{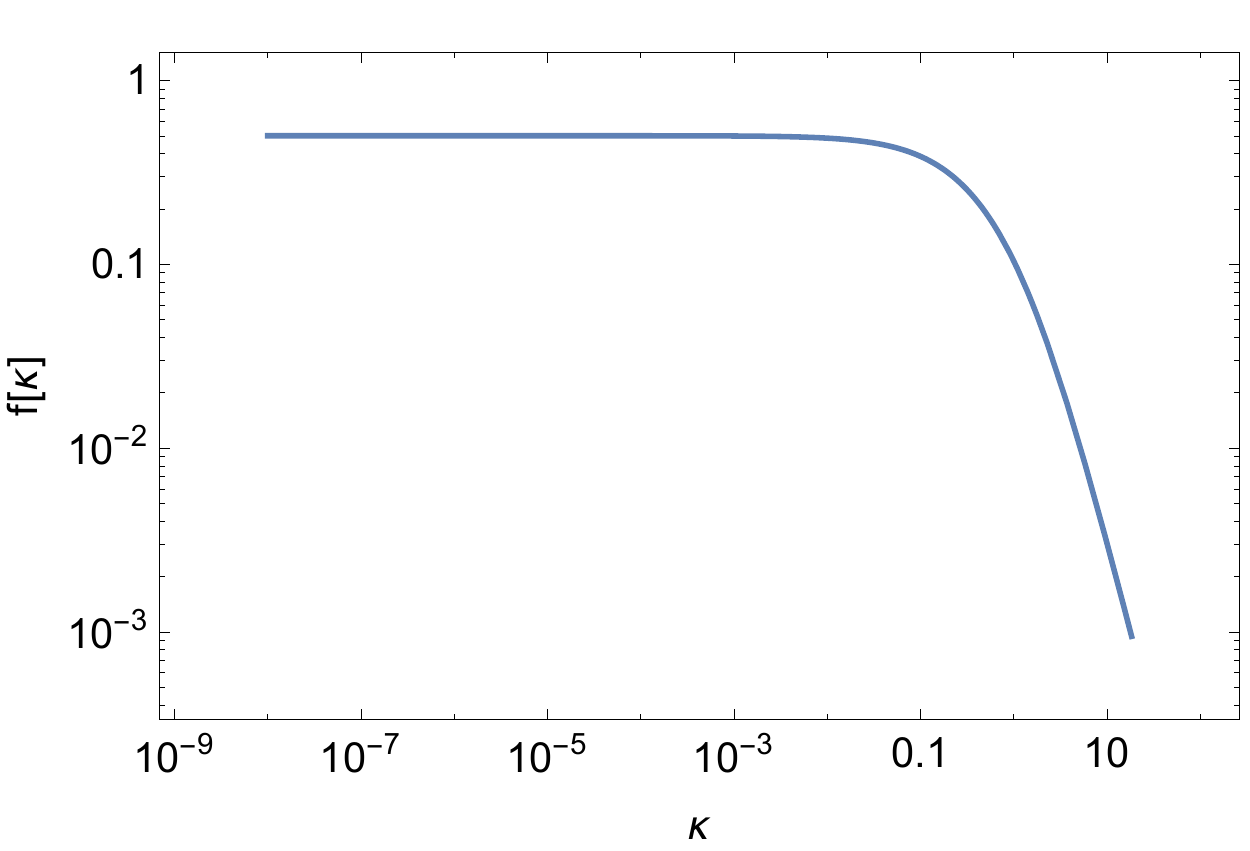}
\label{fig2b}
}
\end{minipage}
\caption{\small{Graph  (a) shows the   behaviour of $f(\kappa)\,\mbox{versus}\,\kappa$ for a wide range of  values of $\kappa$ while graph (b) shows the 
behaviour of $f(\kappa)$ with $\kappa$ in a log-log plot, showing a zone  of fast decrease for $\kappa\sim 2$. This feature can be better appreciated  in Figure 3. }}
\end{figure}
%%%%%%%END FIGURE%%%%%%%%%

The gyromagnetic factors coming from  the two diagrams (a) and (b) ($g_\chi$ and $g_\chi'$)
turn out to be (see \eqref{g} and \eqref{F})
\begin{align} 
g_\chi -2 & = 2\,F_2^{(a)} =\frac{1}{4\pi^{2}}\,f(\kappa), 
\label{222}\\
g_\chi'-2 &= 2\,F_2^{(b)} =\frac{\xi_A}{4\pi^{2}}\, f(\kappa), 
\label{333}
\end{align}
implying  the  identity
\bb
g_\chi'-2 = \xi_A (g_\chi -2).\label{identi}
\ee
This relates the gyromagnetic factor in the dark sector to that for interaction with the visible photon.
\medskip

A final comment is in order here. A different diagram, which can also contribute 
to the gyromagnetic factor of the dark fermion, can be obtained by replacing  
the internal dark photon propagator by that of the visible photon. Although this diagram is allowed from the interactions
in  (\ref{dark3}), it is suppressed by a factor of  ${\xi_A}^2$ (each visible photon vertex would have an extra factor of $\xi_{A}$) which is very small \cite{redondo} and we are discarding terms of this order in this calculation.

%%%%FIGURE 3%%%%%%%%%%%%%%%
\begin{figure}[h!]
\centering
{
\includegraphics[scale=1]{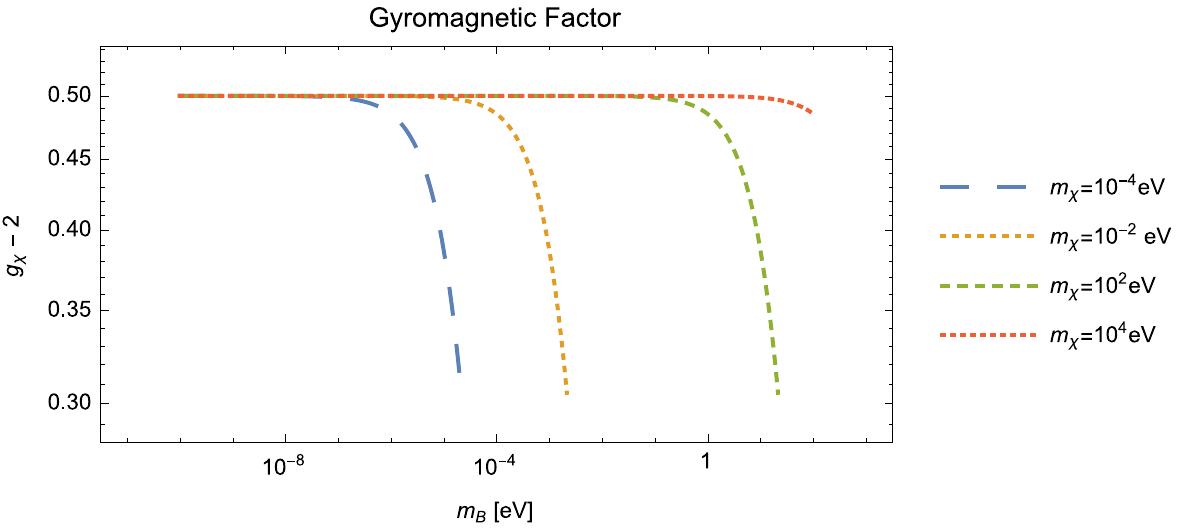}
}
\caption{\label{3a} \small{ This figure shows the values of  $f(\kappa)=g_\chi -2$ as a function 
of the mass of the hidden photon ($m_B$), for different values  
of the dark fermion mass ($m_\chi$), in a log-log plot. }}
\end{figure}
%%%%%%%%%END FIGURE 3%%%%%%%%%%%

Possible values of $g_\chi -2$ factor are shown in Figure \ref{3a} for different values of the fermion masses as a function of the gauge boson mass.  We note from the shape of $f(\kappa)$ in Figure 2 (a) that, as  
the mass of the dark fermion becomes larger (for a fixed value of $m_{B}$), the value of the gyromagnetic factor approaches the 
value of the gyromagnetic factor of the electron only if the electric
charge of the hidden fermion is equal to the electric charge of the electron. 
This behavior is expected because for heavy dark fermions the mass of the hidden photon can be (effectively) neglected which is like the visible sector (zero mass) photon. In this case, one obtains the same gyromagnetic factor provided 
the hidden electric charge is equal to the electron charge. It is also interesting to note that for low mass dark fermions, the value of $g_\chi -2$ 
decreases very fast for fixed  values of $m_B$.  This behavior is generic 
as it can be seen from Fig. \ref{fig2b} \cite{engel, eidelman, redondo}. The $g_\chi'-2$, on the other hand, becomes negative, $g'_{\chi} - 2 <0$, which can be seen from (\ref{identi}) together with the fact that $\xi_{A} <0$
 {and $m_B > m_A$ }(see the sentence following \eqref{dark3})
and this is  different from QED. {In Figure \ref{fig3b} we appreciate  the behavior of $2 - g'_\chi$ for 
$\xi \approx 10^{-16}$. For this value of $\xi$ (and lower values also) we are in the allowed zone of the 
parameter space $(m_B,\xi)$ which is the region colored in gray  in figure  \ref{fig3b} \cite{engel, eidelman, redondo}. We observe that 
the behavior of $2-g'_\chi$ is similar for dark fermions of different masses only for $m_B < 10^{-4}$ or smaller, 
even tough the value of $2-g'_\chi$ is always  tiny. As $m_B \to m_A$, the value of this component of the 
magnetic moment increases as can be seen directly from $\xi_A$ in (\ref{palma}).  }

%%%%%%%%%%%%%%%%%%%%%%%%%%
%%%%%%% F I G U R A 4%%%%%%%%%%%%
%%%%%%%%%%%%%%%%%%%%%%%%%%
\begin{figure}[h!]
\includegraphics[scale=1]{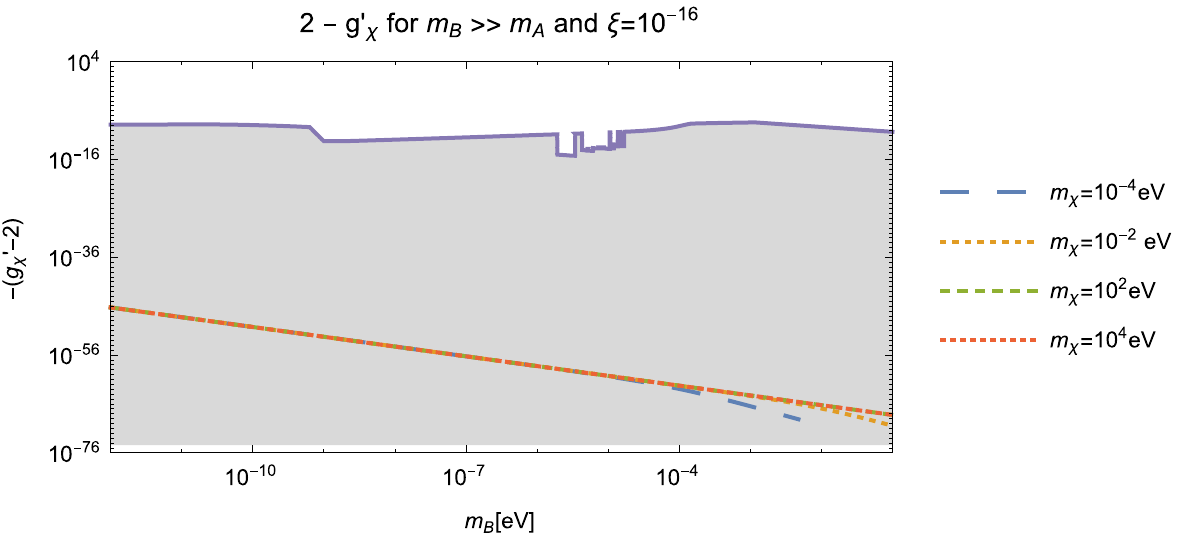}
\caption{\label{fig3b}\small{ Diagram is 
the log-log plot of  the  gyromagnetic factor $g'_\chi-2$ ( that is when dark  fermion couples to visible photons) as a function of $m_B$ for $m_A \approx 10^{-27} $ eV}} 
\end{figure}
%%%%%%%%%END FIGURE 3%%%%%%%%%%%
% (a) shows values of  $g_\chi -2$ as a function 
%of the mass of the hidden photon ($m_B$), for different values of the 
%mass of the dark fermion ($m_\chi$), in a log-log plot. Right diagram

% (see the discussion following \eqref{dark3}) and this is  different from QED. On the other hand, when $m_B\ll m_A$ then $g_\chi'-2 \sim \xi$, as can be seen from (\ref{palma}), and the correction is more significant. Figure (\ref{3a}) shows variations of  $g_\chi-2$ for increasing $m_B$. This plot is completely consistent with the presently known bounds \cite{engel, eidelman, redondo} and, therefore, the greatest contribution to the gyromagnetic factor occurs for dark fermion masses below $10^{-4}~$eV.
%%%%%FIGURE 3%%%%%%%%%%%%%%%
%\begin{figure}[h!]
%\centering
%\begin{minipage}[c]{0.4\textwidth}
%\subfigure[\label{3a}]{
%\includegraphics[scale=0.55]{mg-2NL.pdf}  
%\label{fig3a}
%}
%\end{minipage}
%\begin{minipage}[c]{0.5\textwidth}
%\subfigure[\label{3b}]{
%\includegraphics[scale=0.75]{Fig3A.pdf}
%\label{fig3b}
%}
%\end{minipage}
%\caption{\small{ Diagram (a) shows values of  $g_\chi -2$ as a function 
%of the mass of the hidden photon ($m_B$), for different values of the 
%mass of the dark fermion ($m_\chi$), in a log-log plot. Diagram (b) shows  
%the log-log plot of  the  gyromagnetic factor $g'_\chi-2$ ( that is when dark  fermion couples to visible photons) as a function of $m_B$, for two different values of the mixing parameter $\xi$.}} 
%\end{figure}
%%%%%%%%%%END FIGURE 3%%%%%%%%%%%

\section{Final Comments and Conclusions} 

In this paper we have calculated the magnetic moment of the dark fermion in a model of hidden quantum electrodynamics coupled to visible photons through a kinetic mixing term \cite{Holdom}. 
The kinetic mixing term couples the dark and the visible sectors. The dark fermion has a $g_{\chi}$ (gyromagnetic) factor associated with its interaction with the dark photon as well as a $g'_{\chi}$ factor because of its interaction with the visible photon. An explicit 
relation between $g_\chi$ and $g'_\chi$ has been obtained in (\ref{identi}). From the analysis of 
(\ref{identi}) alongwith comparison with the  data of measurements, we conclude that $g_\chi$ changes 
drastically from the conventional value if the dark fermions are light, otherwise the same behavior, as in standard (visible)  quantum electrodynamics is obtained if the electric charge of the dark fermion is the same as the electron charge.
It is quite possible that in other applications of quantum electrodynamics, such as atomic physics tests and pair production, one may find interesting results.
 
\medskip

This work was supported by FONDECYT/Chile grants 1130020, 1140243 (F.M.),USA-1555 (J.G.)  and  N.T. thanks to the  
Conicyt fellowship 21160064. We would like to  thank Prof. J. E. Kim for useful discussions. 
\section{Appendix: On the Diagonalization Procedure}
\medskip 
In this appendix we would like to give some  details related to the diagonalization of the Lagrangian with the 
kinetic mixing, followed by the mass matrix diagonalization. Let us start with the  Lagrangian density describing
only the  hidden ($B$) and the visible ($A$) gauge fields
 \begin{eqnarray}
{\cal L}& =&  -\frac{1}{4}  F_{\mu  \nu} (A)  F^{\mu  \nu} (A)  -\frac{1}{4}  F_{\mu  \nu} (B)  
F^{\mu  \nu} (B) +
   \frac{\xi}
{2} F_{\mu \nu} (B) F^{\mu \nu} (A) + 
\nonumber
\\
&+&\frac{1}{2} m_B^2 B_\mu B^\mu +\frac{1}{2}{m_A}^2 A_\mu A^\mu.
\nonumber
\\
&\equiv& {\cal L}_g + {\cal L}\big{|}_{\mbox{\tiny{mass}}}.
\label{dark22}
\end{eqnarray}
where ${\cal L}_g$ describes the massless part of ${\cal L}$.

In order to diagonalize the kinetic mixing term in (\ref{dark22}), consider the following linear
transformation of fields
\begin{equation}
\label{diago}
A_{\mu} = {\mathbcal a} \,A'_{\mu} + {\mathbcal b} \,B'_{\mu},\,\,\,\,\,\,\,\,\,\, 
B_{\mu} = {\mathbcal c } \,A'_{\mu} +  {\mathbcal d} \,B'_{\mu}
\end{equation}
with  ${\mathbcal a},  {\mathbcal b}, {\mathbcal c}, {\mathbcal d}\in \Re$. The
massless part of ${\cal L}$  reads now
\begin{eqnarray}
\label{322}
{\cal L}_{g}  =& -&\frac{1}{4}F^{2}(A')\,\left[\ca^2 + \cc^2 - 2\xi\ca\cc\right] - \frac14F^{2}(B') \,
\left[\cb^2 + \cd^2 - 2\xi\cb\cd\right]
\nonumber
\\
&+&\frac{1}{2}F(A')F(B')\,\left[\xi(\ca\cd + \cc\cb) - \ca\cb -\cc\cd\right]
\end{eqnarray}
so that the condition for diagonalization becomes 
\begin{equation}
\label{cond1}
\xi = \frac{\ca\cb +\cc\cd}{\ca\cd + \cc\cb}. 
\end{equation}
This condition reduces the number of parameters in the transformation (\ref{diago}) to three. 
There exist also one-parameter choices of coefficients $\ca,\cb,...$ satisfying (\ref{cond1}) which will 
be discussed at the end of the present section.

We redefine fields  by
$$
\tilde{A}_\mu = A'_\mu\, \sqrt{\ca^2 + \cc^2 - 2\xi\ca\cc},~~~\tilde{B}_\mu = B'_\mu\, \sqrt{\cb^2 + \cd^2 - 2\xi\cb\cd},
$$
and then the Lagrangian density ${\cal L}_g$ turns out to be
\begin{equation}
\label{3}
{\cal L}_{g}  = -\frac14F^{2}(\tilde{A})  - \frac14F^{2}(\tilde{B}).
\end{equation}
The massive part of ${\cal L}$  in (\ref{dark22}) now reads as
\begin{eqnarray}
\label{mass}
{\cal L}\big{|}_{\mbox{\tiny{mass}}} &=& \frac{m_B ^2}{2}\bigg[ \tilde{c}^2(\tilde{A})^2  + \tilde{d}^2 (\tilde{B})^2 +2 (\tilde{A})\cdot (\tilde{B})\,\tilde{c}\,\tilde{d}\bigg] + \frac{{m_A} ^2}{2}\bigg[ \tilde{a}^2(\tilde{A})^2  + \tilde{b}^2 (\tilde{B})^2 +2 (\tilde{A})\cdot (\tilde{B})\,\tilde{a}\,\tilde{b}\bigg],
\end{eqnarray}
with the definitions
\begin{eqnarray}
\label{defini3}
\tilde{a} &=& \frac{\ca}{ \sqrt{\ca^2 + \cc^2 - 2\xi\ca\cc}},~~~~~~~~~~
\tilde{c} = \frac{\cc}{ \sqrt{\ca^2 + \cc^2 - 2\xi\ca\cc}}
\\
\tilde{b} &=&\frac{\cb}{ \sqrt{\cb^2 + \cd^2 - 2\xi\cb\cd}},~~~~~~~~~~
\tilde{d} = \frac{\cd}{ \sqrt{\cb^2 + \cd^2 - 2\xi\cb\cd}}.
\end{eqnarray}â¢
Following \cite{babu}, we rotate fields in order to diagonalize the new mass matrix. 
This rotation does not change the 
kinetic terms and it is given by
\begin{eqnarray}
\label{rot}
{\tilde{A}}_\mu &=& \cos\theta\,A^{''}_\mu + \sin\theta\,B^{''}_\mu,
\\
{\tilde{B}}_\mu &=& -\sin\theta\,A^{''}_\mu + \cos\theta\,B^{''}_\mu.
\end{eqnarray}
The mass term in the Lagrangian can be written now 
\begin{equation}
{\cal L}\big{|}_{\mbox{\tiny{mass}}} = \left(
\begin{array}{cc}
A^{''} & B^{''}
\end{array}\right)_\mu
\,{\mathbbm M}
\left(\begin{array}{c}
{A^{''}}
\\
{B^{''}}
\end{array}\right)^\mu.
\end{equation}
with the mass matrix  ${\mathbbm M}$ given by
\begin{eqnarray}
{\mathbbm M}_{11} &=&(\tilde{c}^2\, m_B^2 +\tilde{a}^2\,{m_A}^2)\cos^2\theta + 
(\tilde{d}^2\, m_B^2 +\tilde{b}^2\,{m_A}^2)\sin^2\theta -
(\tilde{c}\,\tilde{d} \,m_B^2 +\tilde{a}\,\tilde{b}\,{m_A}^2)\sin(2\theta)
\\
{\mathbbm M}_{22} &=&(\tilde{d}^2\, m_B^2 +\tilde{b}^2\,{m_A}^2)\cos^2\theta + 
(\tilde{c}^2\, m_B^2 +\tilde{a}^2\,{m_A}^2)\sin^2\theta +
(\tilde{c}\,\tilde{d} \,m_B^2 +\tilde{a}\,\tilde{b}\,{m_A}^2)\sin(2\theta)
\\
{\mathbbm M}_{12}&=&(\tilde{c}\,\tilde{d} \,m_B^2 +\tilde{a}\,\tilde{b}\,{m_A}^2)\cos(2\theta)  - \frac12\left((\tilde{d}^2-\tilde{c}^2)m_B^2 +
(\tilde{b}^2-\tilde{a}^2){m_A}^2\right)\sin(2\theta) 
\\
&=&{\mathbbm M}_{21}
\end{eqnarray}
The condition for diagonal matrix can be read from here to be 
\begin{equation}
\tan(2\theta) = \frac{2(\tilde{c}\,\tilde{d} \,m_B^2 +\tilde{a}\,\tilde{b}\,{m_A}^2)}
{(\tilde{d}^2-\tilde{c}^2)m_B^2 +
(\tilde{b}^2-\tilde{a}^2){m_A}^2}.
\end{equation}
Finally, we impose the condition for non kinetic mixing (\ref{cond1}) through the 
following parametrization
\begin{equation} 
\ca= \cd =\cosh \tau , ~~~~~~~~~\cb =\cc=\sinh\tau, 
\end{equation}
with the condition $\tanh (2 \tau ) =\xi$. By replacing parameters according to (\ref{defini3}),we find in the limit $\xi\ll 1$
\begin{eqnarray}
{\mathbbm M}_{11}&=& {m_A}^2 -
\frac{{m_A}^4}{m_B^2-{m_A}^2}\,\xi^2+{\cal O}(\xi^4),
\\
{\mathbbm M}_{22}&=&m_B^2+
\frac{m_B^4}{m_B^2-{m_A}^2}\,\xi^2+{\cal O}(\xi^4),
\end{eqnarray}

Finally, we observe that  the vertices in the coupling of hidden fermions 
with the $B$ field are modified since 
\begin{eqnarray}
B_\mu &=& \tilde{c}\tilde{A}_\mu + \tilde{d}\tilde{B}_\mu,
\nonumber
\\
&=& (\tilde{c}\cos\theta -\tilde{d}\sin\theta) A^{''}_\mu + 
(\tilde{c}\sin\theta +\tilde{d}\cos\theta) B^{''}_\mu,
\nonumber
\\
&\equiv& \xi_A\, A^{''}_\mu + \xi_B\, B^{''}_\mu,
\end{eqnarray}
where the last line is a redefinition of the coupling constant. These
coupling constant are related through
\begin{equation}
\xi_A^2+\xi_B^2 = \tilde{c}^2+\tilde{d}^2 =\frac{1}{1-\xi^2},
\end{equation}
which is independent of the parametrization of the restriction. 

The coupling constants modifying the vertices 
in the limit $\xi\ll1$ turn out to be
\begin{eqnarray}
\xi_A&=&-\frac{{m_A}^2}{m_B^2-{m_A}^2}\xi\,+ {\cal O}(\xi^3), \label{palma}
\\
\xi_B &=&
\left(1+\frac{m_B^4 - 2 m_B^2\,{m_A}^2}{2(m_B^2 - {m_A}^2)^2}\,\xi^2\right)+{\cal O}(\xi^3). 
\end{eqnarray}

For  $m_B \gg {m_A}$ we take, for linear corrections in $\xi$
\begin{eqnarray}
\label{couplings}
\xi_A&\approx & -\frac{{m_A}^2}{m_B^2}\xi,
\\
\xi_B &\approx & 1+\xi^2\approx 1.
\end{eqnarray}


\begin{thebibliography}{99}
\bibitem{zeldovich}The influential and important work of Y. Zeldovich and collaborators is summarized in A.~D.~Dolgov and Y.~B.~Zeldovich,
  %`` Cosmology and Elementary Particles,''
  Rev.\ Mod.\ Phys.\  {\bf 53}, 1 (1981).
  %%%%%%%%%%%%%%%%%%%%%%%%%%%%%%%%%%%%%%%%%%%
  \bibitem{kolb} E. Kolb and M. Turner, ``The Early Universe", Addison-Wesley, 1988.
  %%%%%%%%%%%%%%%%%%%%%%%%
\bibitem{review} L.~Goodenough and D.~Hooper,
  %``Possible Evidence For Dark Matter Annihilation In The Inner Milky Way From The Fermi Gamma Ray Space Telescope,''
  arXiv:0910.2998.
  \bibitem{review11} A.~Berlin, P.~Gratia, D.~Hooper and S.~D.~McDermott,
  %``Hidden Sector Dark Matter Models for the Galactic Center Gamma-Ray Excess,''
  Phys.\ Rev.\ D {\bf 90}, 015032 (2014).
  \bibitem{review12} F. Donato, Phys.  Dark Univ. {\bf 4}, 41 (2014); T. Marrod\'an Undagoitia and L. Rauch, J. Phys. G: Nucl. Part. Phys. {\bf 43}, 013001 (2016).
\bibitem{review13}  L. Baudis,  J. Phys. G: Nucl. Part. Phys. {\bf 43}, 044001  (2016). 
%%%%%%%%%%%%%%%%%%%%%%%%%%%%%%%%%%%%%%%%%%%
\bibitem{bertone} G.~Bertone, D.~Hooper and J.~Silk,
  %``Particle dark matter: Evidence, candidates and constraints,''
  Phys.\ Rept.\  {\bf 405}, 279 (2005).
  %%%%%%%%%%%%%%%%%%%%%%%%%%%%%%%%%
  \bibitem{Djouadi} A.~Djouadi,
  %``The Anatomy of electro-weak symmetry breaking. II. The Higgs bosons in the minimal supersymmetric model,''
  Phys.\ Rept.\  {\bf 459}, 1 (2008).
  %%%%%%%%%%%%%%%%%%%%%%%%%%%%%%%%%%%%%%%%%%
\bibitem{test} T.~Kinoshita,
  ``Quantum electrodynamics", World Scientific (1990) 997 p. (Advanced series on directions in high energy physics, 7).
  \bibitem{olive} see also;  K.~A.~Olive {\it et al.} [Particle Data Group Collaboration],
 ``Review of Particle Physics,''
  Chin.\ Phys.\ C {\bf 38}, 090001 (2014).
  %%%%%%%%%%%%%%%%%%%%%%%%%%%%%%%%%%%%%%%%%%%
  \bibitem{babb}  B. ~D\"obrich, \ J. \ Phys. \ Conf. \ Ser. {\bf 632}, 012004 (2015).
  %%%%%%%%%%%%%%%%%%%%%%%%%%%%%%%%%%%%%%%%%%%%
  \bibitem{ry} G. ~Rybka, \ Phys. \ Dark \ Univ. {\bf 4}, 14 (2014);
see also, {\it e.g.}  the proceedings of the 11th Patras Workshop on Axions, WIMPs and WISPs, Zaragoza, June
22 to 26, 2015, https://axion-wimp2015.desy.de/.
  %%%%%%%%%%%%%%%%%%%%%%%%%%%%%%%%%%%%%%%%%%
  \bibitem{ring} J. ~Jaeckel, J.~ Redondo and A. ~Ringwald, \ Phys. \ Rev. \ Lett. {\bf 101}, 131801 (2008).
  %%%%%%%%%%%%%%%%%%%%%%%%%%%%%%%%%%%%%%%%%%%%%%
  \bibitem{coulomb1} D. F. Bartlett, P. E. Goldhagen and E. A. Phillips, \ Phys. \ Rev. {\bf D 2}, 483 (1970).
  %%%%%%%%%%%%%%%%%%%%%%%%%%%%%%%%%%%%%%%%%%%%%%
  \bibitem{coulomb2} E. R. Williams, J. E. Faller and H. A. Hill, \ Phys. \ Rev. \ Lett. {\bf 26}, 721 (1971).
  %%%%%%%%%%%%%%%%%%%%%%%%%%%%%%%%%%%%%%%%%%%%%%%%%%
  \bibitem{planet}  D. ~ Odstrcil, and V.~ J. ~ Pizzo, \ J. \ Geo. Res. {\bf 104}, 28225 (19999.
  
  \bibitem{shift2} J. Jaeckel and S. Roy,  \ Phys. \ Rev. {\bf D 82}, 12020 (2010).
  %%%%%%%%%%%%%%%%%%%%%%%%%%%%%%%%%%%%%%%%%%%%%%%%%%
  
\bibitem{barger} V.~Barger, W.~Y.~Keung and D.~Marfatia, Phys.\ Lett.\ B {\bf 696} (2011) 74
%%%%%%%%%%%%%%%%%%%%%%%%%%%%%%%%%%%%%%%%%%%%%
 \bibitem{pospelov}
  M.~Pospelov and T.~ter Veldhuis, Phys.\ Lett.\ B {\bf 480} (2000) 181.
  %%%%%%%%%%%%%%%%%%%%%%%%%%%%%%%%%%%%%
 \bibitem{kim} W.~S.~Cho, J.~H.~Huh, I.~W.~Kim, J.~E.~Kim and B.~Kyae,
  %``Constraining WIMP magnetic moment from CDMS II experiment,''
  Phys.\ Lett.\ B {\bf 687}, 6 (2010)
  Erratum: [Phys.\ Lett.\ B {\bf 694}, 496 (2011)].
  \bibitem{kim1}  J.~E.~Kim,
  %``Neutrino Magnetic Moment,''
  Phys.\ Rev.\ D {\bf 14}, 3000 (1976).%%%%%%%%%%%%%%%%%%%%%%%%%%%%%%%%%%%%%%%%%%%%%%%%%
  \bibitem{pospelov2}
  M.~Pospelov,   Phys. Rev. D {\bf 80}, 095002 (2009).
  %%%%%%%%%%%%%%%%%%%%%%%%%%%%%%%%%%%%%
  \bibitem{pdg} G. V. Chibisov, \ Usp. \ Fiz. \ Nauk. {\bf 119}, 551 (1976), traslation,  \ Sov. \ Phys. \ Usp.
 {\bf 119}, 624 (1976).
 \bibitem{dolgov} A.~Dolgov and D.~N.~Pelliccia,
  %``Photon mass and electrogenesis,''
  Phys.\ Lett.\ B {\bf 650}, 97 (2007).

\bibitem{barger1} V.~Barger, W.~Y.~Keung, D.~Marfatia and P.~Y.~Tseng,
  %``Dipole Moment Dark Matter at the LHC,''
  Phys.\ Lett.\ B {\bf 717}, 219 (2012)
  
    \bibitem{foadi}   R.~Foadi, M.~T.~Frandsen and F.~Sannino,
  %``Technicolor Dark Matter,''
  Phys.\ Rev.\ D {\bf 80}, 037702 (2009).
  \bibitem{paolo} P.~Gondolo and K.~Kadota,
  %``Late Kinetic Decoupling of Light Magnetic Dipole Dark Matter,''
  JCAP {\bf 1606}, 012 (2016).  
%%%%%%%%%%%%%%%%%%%%%%%%%%%%%%%%%%%%%%%%%%%%%
  \bibitem{Holdom} B. Holdom,  Phys.\ Lett.\ B {\bf 166} (1986) 196.
  %%%%%%%%%%%%%%%%%%%%%%%%%%%%%%%%%%%%%%%
\bibitem{engel} See for example, 
  D.~Veberic {\it et al.} [FUNK Experiment Collaboration],  PoS ICRC {\bf 2015}, 1191 (2015).
 %%%%%%%%%%%%%%%%%%%%%%%%%%%%%%%%%%% 
   \bibitem{monton}   J.~L.~Feng, J.~Smolinsky and P.~Tanedo,
  Phys.\ Rev.\ D {\bf 93} (2016),  115036.
  %%%%%%%%%%%%%%%%%%%%%%%%%%%%%%%%%%%%%%%
    %%%%%%%%%%%%%%%%%%%%%%%%%%%%%%%%%%%%%%%
  \bibitem{schwinger} 
  J.~S.~Schwinger,
  %``On Quantum electrodynamics and the magnetic moment of the electron,''
  Phys.\ Rev.\  {\bf 73}, 416 (1948).  
    %%%%%%%%%%%%%%%%%%%%%%%%%%%%%
\bibitem{eidelman}
  K.~A.~Olive {\it et al.} [Particle Data Group Collaboration],
  ``Review of Particle Physics,''
  Chin.\ Phys.\ C {\bf 38} (2014) 090001.
  %%%%%%%%%%%%%%%%%%%%%%%%%%%%%%%%%%%%%%%
  \bibitem{redondo} For a review see, J.~Redondo,
  %``Atlas of solar hidden photon emission,''
  JCAP {\bf 1507}, no. 07, 024 (2015).
%  \bibitem{helio} For a review, see for example  R.~Essig {\it et al.},
%  %``Working Group Report: New Light Weakly Coupled Particles,''
%  arXiv:1311.0029 [hep-ph].  
  \bibitem{babu}  K.~S.~Babu, C.~F.~Kolda and J.~March-Russell,
  %``Leptophobic U(1) $s$ and the R($b$) - R($c$) crisis,''
  Phys.\ Rev.\ D {\bf 54} (1996) 4635; ibid, 
  %``Implications of generalized Z - Z-prime mixing,''
  Phys.\ Rev.\ D {\bf 57} (1998) 6788  
 \end{thebibliography}
\end{document}